\documentclass[nofootinbib,twocolumn,prd,preprintnumbers,superscriptaddress,aps]{revtex4-1}

\usepackage{amsfonts}
\usepackage{graphicx}
\usepackage[colorlinks=true,linkcolor=blue,citecolor=teal,urlcolor=blue]{hyperref}
\hypersetup{breaklinks=true}
\usepackage{xcolor}
\usepackage{amsmath}
\usepackage{cases}
\usepackage{braket}
\usepackage[T1]{fontenc}
\usepackage{mathrsfs}
\usepackage{enumerate}
\usepackage{bm}
\usepackage{multirow}
\usepackage{comment}
\usepackage{mathrsfs}

\begin{document}

\title{Lagrangian formulation of the Tsallis entropy}

\author{Rocco D'Agostino}
\email{rocco.dagostino@unina.it}
\affiliation{Scuola Superiore Meridionale, Largo San Marcellino 10, 80138 Napoli, Italy}
\affiliation{Istituto Nazionale di Fisica Nucleare (INFN), Sezione di Napoli, Via Cinthia 21, 80126 Napoli, Italy}

\author{Giuseppe Gaetano Luciano}
\email{giuseppegaetano.luciano@udl.cat}
\affiliation{Department of Chemistry, Physics and Environmental and Soil Sciences, Escola Politecninca Superior, Universidad de Lleida, Av. Jaume II, 69, 25001 Lleida, Spain}

\begin{abstract}
We investigate the gravitational origin of the Tsallis entropy, characterized by the nonadditive index $\delta$. Utilizing Wald's formalism within the framework of $f(R)$ modified theories of gravity, we evaluate the entropy on the black hole horizon for constant curvature solutions to the spherically symmetric vacuum field equations. In so doing,  
we demonstrate that the Tsallis entropy can be effectively derived from a generalization of the Lagrangian $\mathcal{L}\propto R^{1+\epsilon}$, where $\epsilon=\delta-1$ quantifies small deviations from general relativity.
In conclusion, we examine the physical implications of our findings in light of cosmological observations and elaborate on the possibility of solving the thermodynamic instability of Schwarzschild black holes. 
\end{abstract}

\maketitle

\section{Introduction}
The qualitative and quantitative success of Boltzmann-Gibbs (BG) statistical mechanics makes it one of the most robust theoretical frameworks of physics. 
Applications are diverse and nearly ubiquitous, including equilibrium thermodynamics, high-energy physics, information theory and solid state physics, at both classical and quantum levels \cite{landau2013statistical}.
Despite such enormous resonance,  
a more comprehensive examination uncovers specific limitations and inadequacies for certain complex systems, highlighting the lack of universality of some of its foundational principles. Concretely, it was noted by Gibbs \cite{gibbs1902elementary} that systems with a divergent partition function, such as large-scale gravitational systems, cannot be explained by BG statistics. The long-standing puzzle surrounding the statistical mechanics of self-gravitating systems provides an illustrative example of this shortcoming \cite{1957SvA.....1..748O,10.1093/mnras/136.1.101}.
In a similar vein, the computation of black hole (BH) entropy by the standard BG machinery leads to the well-known \emph{area law} \cite{Bekenstein:1973ur,Hawking:1974rv}, which dramatically violates the extensivity of thermodynamic entropy if BHs are to be considered genuine $(3+1)$-dimensional systems \cite{tHooft:1984kcu,Susskind:1993ki,Srednicki:1993im,Strominger:1996sh}.
These limits have naturally prompted exploration into generalized formalisms that could enlarge the realm of applicability of the standard statistical mechanics and thermodynamics.

A promising framework in the quest for generalized statistics is the Tsallis theory \cite{Tsallis:1987eu}, which is built upon a \emph{nonadditive} entropic functional incorporating the BG limit as a special case. The rationale behind the Tsallis paradigm is that the fundamental hypothesis of the BG entropy, namely the weak probabilistic correlations between the elements of a given system, fails in the presence of long-range interactions and/or strong quantum entanglement. This breakdown motivates  
a suitable redefinition of entropy that takes into account the above complexities. So far, Tsallis statistics has proven successful in describing a wide and important class of physical systems characterized by either non-ergodicity, long-term memories, multi-fractal nature or dissipation \cite{e13101765,Lyra:1997ggy,Luciano:2021onl,Luciano:2021mto}. Interestingly, growing attention has recently been dedicated to cosmological implications. Using the gravity-thermodynamic conjecture~\cite{Jacobson:1995ab} with the Tsallis entropy, modified Friedmann equations have been derived from the application of the first law of thermodynamics to the Universe horizon \cite{Abreu:2012msk,Sheykhi:2018dpn,Lymperis:2018iuz}. Dark energy models through nonextensive statistics have been explored in Refs.~\cite{Barboza:2014yfe,Tavayef:2018xwx,SayahianJahromi:2018irq,Saridakis:2018unr,DAgostino:2019wko,Luciano:2022ffn}.
Additionally, interesting cosmological phenomenology was analyzed in Refs.~\cite{Nojiri:2019skr,Zamora:2022cqz,Luciano:2022ely,Nojiri:2022dkr,Jizba:2022bfz,Basilakos:2023kvk,Teimoori:2023hpv}, while connections with Planck-scale deformations of the Heisenberg uncertainty relation appear in Refs.~\cite{Shababi:2020evc,Luciano:2021ndh,Jizba:2022icu,Jizba:2023ygi}. Besides the Tsallis formulation, we mention that several other nonextensive entropies are commonly used in the literature, motivated by either information theory \cite{Reny,sharma1975new}, relativity \cite{Kaniadakis:2002zz,Luciano:2022eio} or quantum gravity \cite{Barrow:2020tzx}.

While the implications of the Tsallis generalized entropy have been largely considered for BH systems \cite{Tsallis:2012js}, the corresponding modified theory of gravity has not yet been explored at a fundamental level. 
This may somehow challenge the significance of the Tsallis entropy in enhancing our understanding of the Universe. 
In fact,  modifications to General Relativity (GR) have been extensively explored over the last years as a possible solution to cosmological issues, such as vacuum energy and dark matter problems \cite{Bengochea:2008gz,Clifton:2011jh,DAgostino:2018ngy,Capozziello:2019cav,Deser:2007jk,DAgostino:2019hvh,Koyama:2015vza,CANTATA:2021ktz,DAgostino:2022tdk,Nojiri:2017ncd,Califano:2023aji,Li:2007jm,DAgostino:2024ojs}. Relevant examples in this regard are the $f(R)$ theories \cite{Carroll:2003wy,Capozziello:2003tk,Starobinsky:2007hu,Sotiriou:2008rp,DeFelice:2010aj}, where the gravitational Lagrangian is generalized to an arbitrary function of the Ricci scalar, $R$.

In light of the above considerations, this work aims to investigate the roots of the Tsallis BH entropy in gravitational theory. Preliminary efforts to accommodate Tsallis corrections in the gravitational action have been conducted in different cosmological contexts through various reconstruction schemes \cite{Ghaffari:2018wks,Aditya:2019bbk,Waheed:2020cxw,DiGennaro:2022grw,Luciano:2023wtx}. 
In this paper, we intend to use the Noether charge method based on Wald's approach \cite{Wald:1993nt,Iyer:1994ys} to infer, at a more fundamental level, the gravitational Lagrangian giving rise to the modified Tsallis entropy. According to this formalism, the entropy of BH's event horizon can be found within any diffeomorphism invariant theory of gravity. It is possible to verify that Wald's formula consistently reproduces the standard area law in GR, while corrections do arise for any other gravitational model that deviates from the Einstein-Hilbert action. Hence, if a given fundamental approach predicts any departure from the Bekenstein-Hawking entropy, one can derive the corresponding modified Lagrangian by profitably applying Wald's method the other way around (see also Ref.~\cite{Hammad:2015dka}). We here use this strategy to show that the Tsallis entropy originates from a special class of $f(R)$ gravity theories.

The structure of the paper is as follows. In Sec.~\ref{Tsallis} we discuss the mathematical background of the Tsallis and Wald formulations of entropy. In Sec.~\ref{sec:f(R)_BHs}, we briefly review the spherically symmetric solutions in $f(R)$ theories of gravity. Within this framework, we reconstruct the effective Tsallis Lagrangian starting from corrections to the Schwarzschild BH metric. Implications for cosmology and thermodynamic stability of BHs are discussed in Sec.~\ref{PhysImpl}, while conclusions and outlook are finally summarized in Sec.~\ref{Concl}. Throughout this paper, we use the spacetime signature $(-,+,+,+)$. We also adopt units such that $c=k_B=\hbar=1$, unless otherwise explicitly specified.

\section{Tsallis and Wald entropic formalisms}
\label{Tsallis}

Entropy is a nuanced yet subtle concept in physics. Despite being introduced as a property in thermodynamics, it notoriously plays a role in a variety of contexts from statistical mechanics, to gravity and information theory. A crucial implication in quantum gravitational applications of BG entropy is holography, which states that the description of a volume of space can be thought of as encoded on a lower-dimensional boundary to the region~\cite{tHooft:1993dmi,Susskind:1994vu}. However, in the early 1900s, Gibbs highlighted that the BG theory is inapplicable to systems where the partition function diverges, and we know that gravitational systems lie
within this class. Later on, Tsallis introduced a thermodynamic entropy that is suitable for systems with a sub-extensive scaling of microstates, such as BHs or certain quantum condensed-matter systems~\cite{Tsallis:1987eu}. The resulting entropic functional generalizes the BG definition according to
\begin{equation}
\label{TsEn}
    S_\delta=\sum_{i=1}^W p_i\left(\ln\frac{1}{p_i}\right)^\delta\,,\quad \delta>0\,,
\end{equation}
where $p_i$ denotes the probability of the $i$-th microstate of the system and $W$ is the total number of microstates. 
In the case of equiprobable states, it follows that
\begin{equation}
\label{TsEn2}
    S_\delta=\left(\ln W\right)^\delta\,.
\end{equation}

To check the consistency of the Tsallis paradigm, let us consider the emblematic case of $(3+1)$-dimensional BHs. In the standard BG framework, the corresponding entropy obeys the well-known area-law 
\begin{equation}
\label{arealaw}
S_\text{BG}=-\sum_{i=1}^W p_i\ln p_i\propto L^2\,,  
\end{equation}
where $L$ is the characteristic size of the system. 
Using the asymptotic equipartition property, we have $S_\text{BG}=\ln W$, which entails~\cite{Tsallis:2012js}
\begin{equation}
\label{WL}
    W(L)\propto\phi(L)\,\xi^{L^2}\,,\quad \xi>1
\end{equation}
for any function $\phi(L)$ satisfying 
\begin{equation}
    \lim_{L\rightarrow\infty}\frac{\ln\phi(L)}{L^\gamma}=0\,, \quad 0<\gamma<1\,.
\end{equation}
It is easy to verify that, while the behavior~\eqref{WL} prevents Eq.~\eqref{arealaw} from being identified with the correct (i.e. extensive) thermodynamic entropy, Eq.~\eqref{TsEn2} gives the desired scaling for $\delta=3/2$. Moreover, $S_\delta$  preserves the Legendre-transform structure of thermodynamics~\cite{e22010017}. 

In BH mechanics (and, by extension, in cosmology~\cite{Gibbons:1977mu}) the area-law scaling~\eqref{arealaw} is associated with the surface area $\mathcal{A}$ of the BH horizon. In this case, from Eqs.~\eqref{TsEn2} and~\eqref{WL}, we can write the  Tsallis entropy in the equivalent form~\cite{Tsallis:2012js} 
\begin{equation}
    S_\delta=
    \Gamma_\delta{\mathcal{A}}^\delta\,.
    \label{eq:Tsallis entropy}
\end{equation}
As discussed earlier, under the hypothesis of equal probabilities, the parameters $\Gamma_\delta$ and $\delta$ are related to the dimensionality of the system. Nevertheless, in the general case, they remain as completely free parameters. For concreteness, we can set $\Gamma_\delta=\alpha^{1-\delta}G^{-\delta}/4$, where $\alpha>0$ is a pure number. Note that, by this definition, $S_\delta$ is correctly dimensionless (in the adopted units) for any $\delta$. Furthermore, the standard Bekenstein-Hawking expression is recovered when $\delta=1$.

\subsection{Wald’s method}
To construct the Lagrangian theory of gravity consistent with Tsallis entropy~\eqref{eq:Tsallis entropy}, we now invoke Wald's formula. In his seminal work \cite{Wald:1993nt}, Wald demonstrated that the BH entropy is the Noether charge associated with any
diffeomorphism invariant theory of gravity that admits stationary solutions with a bifurcate Killing horizon. When applied to the Einstein-Hilbert action of GR, such an approach yields the ordinary Bekenstein-Hawking entropy, while it introduces an extra term depending on the curvature of the background spacetime in the context of extended theories of gravity \cite{Vollick:2007fh,Briscese:2007cd,delaCruz-Dombriz:2009pzc}. Conversely, given an entropy that deviates from the  Bekenstein-Hawking horizon scaling, one might take advantage of Wald's formula to extract the underlying modified gravity model. 

To concretely apply Wald's method to the Tsallis entropy~\eqref{eq:Tsallis entropy}, let us briefly discuss the derivation of Wald's formula as originally proposed in \cite{Wald:1993nt,Iyer:1994ys}. Given a theory of gravity described by Lagrangian $\mathcal {L}(\phi)$, where $\phi$ denotes the field
variables which the theory depends on, it is diffeomorphism invariant iff $\mathcal L$ remains unchanged under arbitrary spacetime transformations. According to Noether's theorem, one can define a Noether three-form current $\textbf{j}$ and (for solutions to the field equations) a two-form charge $\textbf{Q}$, such that $\textbf{j}=d\textbf{Q}$. The explicit expression of $\textbf{j}$ can be found by noticing that, under the variation of the dynamical fields, the variation of the four-form $\textbf{L}=\mathcal{L}(\phi)dx^0\wedge...\wedge dx^3$ is in the form $\delta\textbf{L}=\textbf{E}\,\delta\phi+d{\bf\Theta}(\phi,\delta\phi)$. In turn, this gives the dynamical equations $\textbf{E}$ (in form notation) up to the total divergence of the three-form ${\bf{\Theta}}$. The associated Noether current is then 
$\textbf{j}={\bf\Theta}(\phi,\mathcal{L}_\xi\phi)-\xi\cdot{\bf{L}}=d\textbf{Q}$,
where $\mathcal{L}_\xi$ denotes the Lie is with respect to the vector field
$\xi=\xi^\mu\partial_\mu$ generating the diffeomorphism, and ``$\cdot$'' denotes the contraction of a vector field with the first index of a differential form. 

Denoting by $\mathcal{H}$ the bifurcation surface
of a BH horizon, the flux of the current $\textbf{j}$ through $\mathcal{H}$ corresponds to the entropy transfer between two
spacetime regions separated by $\mathcal{H}$. Then, the entropy of a BH is given by the surface integral \cite{Wald:1993nt,Iyer:1994ys}
\begin{equation}
    S=\frac{2\pi}{\kappa}\int_\mathcal{H}\textbf{Q}\,,
\end{equation}
with $\kappa$ being the surface gravity of the BH. In the case of a generic gravitational theory $\mathcal{L}(g_{\mu\nu},R_{\mu\nu\rho\sigma},\nabla_\mu R_{\nu\rho\sigma\lambda,...})$, identifying $\xi$ with the horizon Killing vector\footnote{Notice that $\chi^\mu=0$ on the surface $\mathcal{H}$.} $\chi$, Wald's entropy reads
\begin{equation}
\label{Swald}
S=-2\pi\int_\mathcal{H} d\mathcal{A}\, \varepsilon_{\mu\nu}\varepsilon_{\rho\sigma}\, \frac{\delta\mathcal{L}}{\delta R_{\mu\nu\rho\sigma}}\,,
\end{equation}
where $R_{\mu\nu\rho\sigma}$ is the Riemann tensor, while $\varepsilon_{\mu\nu}\equiv \nabla_\mu \chi_\nu $ is the binormal vector of the horizon such that $\varepsilon_{\mu\nu}=-\varepsilon_{\nu\mu}$ and $\varepsilon_{\mu\nu}\varepsilon^{\mu\nu}=-2$. Here, the functional derivative is carried out at fixed metric $g_{\mu\nu}$ and connection $\Gamma_{\mu\nu}^\rho$.

\section{Tsallis Lagrangian from $f(R)$ gravity}
\label{sec:f(R)_BHs}

Wald's entropy formula \eqref{Swald} can be applied to specific choices of the Lagrangian density. In what follows, we consider the case of $f(R)$ theories of gravity with action \cite{Sotiriou:2008rp,DeFelice:2009rw} 
\begin{equation}
    \mathscr{S}=\dfrac{1}{16\pi G}\int d^4x\, \sqrt{-g}\, f(R)\,.
    \label{eq:f(R) action}
\end{equation}
Here, $G$ is the Newton constant, $f(R)$ is generic function of the Ricci curvature, $R=g_{\mu\nu}R^{\mu\nu}$, $R_{\mu\nu}$ is the Ricci tensor and $g$ is the determinant of the metric tensor $g_{\mu\nu}$.

The variation of the above action with respect to $g_{\mu\nu}$ provides us with the field equations
\begin{equation}
    f'(R)R_{\mu\nu}-\dfrac{1}{2}g_{\mu\nu}f(R)-(\nabla_\mu\nabla_\nu-g_{\mu\nu}\Box)f'(R)=0\,,
    \label{eq:FE}
\end{equation}
where the prime indicates the first derivative with respect to $R$, and $\Box\equiv g_{\mu\nu}\nabla^\mu\nabla^\nu$ is the D'Alembert operator. It is worth noting that Eq.~\eqref{eq:FE} reduces to Einstein's equations of GR as soon as $f(R)=R$.
An additional differential equation can be obtained by taking the trace of the field equations:
\begin{equation}
    f'(R)R-2f(R)+3\Box f'(R)=0\,,
    \label{eq:trace}
\end{equation}
where the `0' on the right-hand side is replaced by the trace of the stress-energy tensor when matter is present.

To study the spherically symmetric solution in $f(R)$ gravity, we consider the spacetime line element 
\begin{equation}
    ds^2=-h(r)dt^2+\dfrac{1}{h(r)}dr^2+r^2\left(d\theta^2+\sin^2\theta\, d\phi^2\right),
    \label{eq:metric}
\end{equation}
where the metric function $h(r)$ accounts for the particular BH solution. Specifically, one might look for constant curvature solutions $R=R_0\in\mathbb{R}$ satisfying \cite{Multamaki:2006zb}
\begin{equation}
    f'(R_0)R_0 -2f(R_0)=0\,.
    \label{eq:const_sol}
\end{equation}
In this case, one finds 
\begin{equation}
\label{hsol}
    h(r)=1-\frac{2GM}{r}-\dfrac{R_0 r^2}{12}\,.
\end{equation}
This metric function represents the Schwarzschild-de Sitter BH solution arising from the Lagrangian density $f(R)=R-2\Lambda$, with $\Lambda>0$ being the cosmological constant. The latter is related to $R_0$ through the relation $R_0=4\Lambda$.

The condition $h(r)=0$ provides us with the horizon radius, $r_\mathcal{H}$. In our case, we can perturb around the Schwarzschild solution to obtain
\begin{equation}
\label{rh}
    r_\mathcal H= r_S\left(1+\frac{R_0 }{12}r_S^2\right)+\mathcal{O}(R_0^2)\,,
\end{equation}
where $r_S\equiv 2G M$ is the Schwarzschild radius that is recovered for $R_0\rightarrow0$. Accordingly, the area of the event horizon is given by
\begin{equation}
    \mathcal{A}=4\pi r_\mathcal{H}^2=4 \pi r_S^2 \left(1+\frac{R_0}{6} r_S^2\right)+\,\mathcal{O}(R_0^2)\,.
    \label{eq:area}
\end{equation}

Let us now notice that, for gravitational theories like Eq.~\eqref{eq:f(R) action}, Wald's entropy~\eqref{Swald} can be cast as~\cite{Vollick:2007fh,Briscese:2007cd}
\begin{equation}
    S=\dfrac{\mathcal{A}}{4G}\left[f'(R)\right]_{\mathcal{H}}\,.
\end{equation}
For our purposes, we need to evaluate $f'(R)$ on the event horizon, $\mathcal{H}$. This becomes particularly simple when considering a constant curvature solution. In fact, Eq.~\eqref{eq:const_sol} admits two classes of solution, namely $f(R)\propto R^2$ and $R=$ const. In the latter case, assuming $R=R_0$ everywhere, one has
\begin{equation}
    S=\dfrac{\mathcal{A}}{4G}\left[f'(R)\right]_{R=R_0}\,.
    \label{eq:S_formula}
\end{equation}
Comparing this with Eq.~\eqref{eq:Tsallis entropy} and using Eq.~\eqref{eq:area}, we obtain
\begin{equation}
    f'(R)= \left[\frac{16 \pi  G M^2}{\alpha} \left(1+\frac{2}{3} G^2 M^2 R\right)\right]^{\delta -1} \,,
    \label{eq:f'(R)}
\end{equation}
which can be finally integrated over $R$ to yield
\begin{equation}
    f(R)=c_1+\frac{3}{2 G^2 M^2 \delta}\left(\frac{16\pi GM^2}{\alpha}\right)^{\delta-1}\left(1+\frac{2}{3} G^2 M^2 R\right)^{\delta}.
    \label{eq:f(R)}
\end{equation}
The arbitrary constant $c_1$ can be fixed by evaluating Eq.~\eqref{eq:f(R)} in the standard entropy scenario:
\begin{equation}
   f(R)\Big|_{\delta=1}=c_1+R+\frac{3}{2 G^2 M^2}\,.
\end{equation}
The above expression recovers the Schwarzschild-de Sitter solution, corresponding to $f(R)=R-2\Lambda$, if 
\begin{equation}
    c_1=-\frac{3}{2 G^2 M^2}-2 \Lambda\,.
    \label{eq:c_1}
\end{equation}
In the case of vanishing cosmological constant and $\delta=1$, the obtained solution corresponds to the Hilbert-Einstein Lagrangian leading to the Schwarzschild spacetime.

Now, we shall look for an effective Lagrangian that can mimic Eq.~\eqref{eq:f(R)} in the perturbation regime of the Bekeinstein-Hawking entropy.  
We then compute the first-order Taylor series around $\delta=1$ to obtain\footnote{ Hereafter, we denote by `$\approx$' quantities expanded to first order around both $R_0=0$ and $\delta=1$.}
\begin{align}
    &f(R)\approx c_1 +\left(\frac{3}{2 G^2 M^2}+R\right)\times \nonumber \\
    &\times \Bigg\{1+(1-\delta )\left\{1-\ln\left[\frac{16 \pi  G M^2}{\alpha} \Big(1+\frac{2}{3} G^2 M^2 R\Big)\right]\right\}\Bigg\}.
    \label{eq:multiv_Taylor}
\end{align}
{We notice that the latter is equivalent to the first-order expansion around $\delta=1$ of the function
\begin{equation}
    f(R)= c_1+c_2(1+c_3 R)\left[1+(\delta-1)\ln (1+c_3 R)\right],
    \label{compactfR}
\end{equation}
where we have defined
\begin{subequations}
    \begin{align}
\label{c2}
    c_2&\equiv \frac{1}{c_3}\left\{1+(1-\delta)\left[1-\ln\left(\frac{16 \pi  G M^2}{\alpha }\right)\right]\right\}, \\
    \label{c3}
    c_3&\equiv\frac{2}{3}G^2M^2\,.
    \end{align}
\end{subequations}
In turn, Eq.~\eqref{compactfR} coincides with the first-order expansion around $\epsilon=0$  of the function
\begin{equation}
  f(R)=c_1 +c_2\left(1 + c_3 R\right)^{1+\epsilon}\,,
   \label{eq:f(R)_final}
\end{equation}
where 
\begin{equation}
\label{neweps}
  \epsilon\equiv\delta -1\,.  
\end{equation}
}
Therefore, we can conclude that the Tsallis entropy can be effectively derived from the modified gravity Lagrangian given by Eq.~\eqref{eq:f(R)_final}. 
Notice that $\epsilon=0$ (i.e. $\delta=1$) implies $c_2=1/c_3$. Then, by use of Eq.~\eqref{eq:c_1}, we recover the standard $f(R)=R-2\Lambda$ gravitational model.

As a consistency check, we can plug Eq.~\eqref{eq:f(R)_final} into Eq.~\eqref{eq:S_formula} by means of the definitions~\eqref{c2}, \eqref{c3} and \eqref{neweps}. Then, expanding at the first-order around $\delta=1$ yields
\begin{align}
    S=&\,4 \pi  G M^2 \left(1+\frac{2}{3} G^2 M^2 R_0\right) \nonumber \\
    & \times \left\{1+(\delta -1) \ln \left[\frac{16 \pi  G M^2}{\alpha} \Big(1+\frac{2}{3} G^2 M^2 R_0\Big)\right]\right\}.
\end{align}
The latter can be implicitly written as a function of the horizon area as
\begin{equation}
    S=\frac{\mathcal{A}}{4G}\left[1 + \left(\delta-1\right)\ln \left(\frac{\mathcal{A}}{\alpha\, G}\right)\right].
    \label{eq:S_new}
\end{equation}
It is easy to verify that Eq.~\eqref{eq:S_new} coincides with the first-order Taylor series of the Tsallis entropy~\eqref{eq:Tsallis entropy}. Hence, this proves the goodness of our method and the validity of result \eqref{eq:f(R)_final}.

It is worth mentioning that logarithmic-like corrections to the horizon scaling are commonly encountered in quantum approaches to gravity~\cite{Log1,Log2,Log3,Log4}. In this respect, our treatment represents a further step toward a deeper understanding of the nature of gravity and spacetime.

\section{Physical consequences}
\label{PhysImpl}

The modified gravity Lagrangian~\eqref{eq:f(R)_final} encodes corrections to the Einstein-Hilbert action controlled by the small parameter $\epsilon$. Interestingly, our result may be seen as a generalization of the model $f(R)\propto R^{1+\epsilon}$, whose cosmological and weak-field properties have been first studied in Ref.~\cite{Clifton:2005aj}. Further applications have been later investigated in the contexts of gravitational waves~\cite{Capozziello:2007vd} and BH physics~\cite{DeFalco:2023rxm,DAgostino:2024ymo}. In the following, we consider the implications of our result in cosmology, quantum gravity phenomenology and BH physics. 

\subsection{Cosmology and quantum gravity implications}

The value of $\epsilon$ has been constrained using observational abundances of the light elements~\cite{Clifton:2005aj}. In so doing, the condition $-0.017<\epsilon<0.0012$ has been set working within a suitably generalized cosmology. From the relation~\eqref{neweps}, such a result turns into the $\delta$-constraint
\begin{equation}
\label{CliftBar}
0.983\lesssim\delta\lesssim1.001\,. 
\end{equation}
Remarkably, this condition is consistent with the $2\sigma$ constraints on the Tsallis cosmology obtained from low and high redshift observations~\cite{Asghari:2021lzu}. 

From a more qualitative perspective, we observe that all the aforementioned considerations on the Tsallis model can be directly extended to the Barrow $\Delta$-entropy~\cite{Barrow:2020tzx} upon the identification $\delta\rightarrow1+\Delta/2$, with $0\leq \Delta\leq 1$. Besides this formal similarity, however, the two entropies have conceptually different foundations. On the one hand, the Tsallis paradigm emerges from the classical requirement of extensivity of BH thermodynamics. By contrast, Barrow's entropy incorporates an effective parameterization of quantum fluctuations upon the surface geometry of BHs, which can cause a modification of the topology of spacetime approaching the Planck scale. In light of this observation, we infer that the phenomenology of the extended theory~\eqref{eq:f(R)_final} could be studied to provide preliminary insights into the formulation of a quantum theory of gravity.

\subsection{Thermodynamic stability of black holes}

Within the context of BH thermodynamics, special attention is devoted to the concept of stability, which critically relies on the use of horizon thermodynamic variables
\cite{Davies:1978zz}. 
In this regard, the study of the BH local stability is strictly related to the positivity of the heat capacity, $C$.  Moreover, singularities or sign changes of $C$ signal the occurrence of a phase transition and the breakdown of the equilibrium thermodynamic description. 

In standard GR, the Schwarzschild solution leads to BHs that are locally unstable. 
{To see this, we remind that Hawking's temperature can be derived from
\begin{equation}
    T_H=\left(\partial_M S\right)^{-1}\,,
\end{equation}
which gives for the heat capacity~\cite{Czinner:2015eyk}
\begin{equation}
\label{HC}
   \mathcal{C}=-\frac{(\partial_M S)^2}{\partial_M^2 S}\,.
\end{equation}}
Using the Bekenstein-Hawking area law $S=A/(4G)$, after some simple algebra, we obtain $C=-8\pi G M^2<0$. The negative sign of $C$ defines the local thermodynamic instability of the Schwarzschild BH. Phenomenologically, this corresponds to the fact that the BH heats up as it radiates energy. {Nevertheless, it is worth noting that the Schwarzschild BH can
be stabilized by a cavity confinement~\cite{York:1986it} or in the presence of a sufficiently large cosmological constant~\cite{Hawking:1982dh} while remaining within the standard BG thermodynamics.

We shall now inspect the aforementioned considerations in light of Tsallis's paradigm. 
For this purpose, we rewrite Eq.~\eqref{eq:Tsallis entropy} as 
\begin{equation}
\label{NewTs}
       S_\delta =
       \frac{1}{4} \left(\frac{4 \pi  r_{\mathcal{H}}^2}{G}\right)^{\delta }\,,
\end{equation}
where we have set the free parameter $\alpha$ to unity for the sake of simplicity.
To obtain the heat capacity for the Tsallis entropy given by Eq.~\eqref{NewTs}, we shall use the horizon radius as in Eq.~\eqref{rh}, which has been derived in the perturbation regime around the Schwarzschild BH solution. Therefore, for consistency, we must take into account only the leading order in $R_0$, so obtaining
\begin{align}
\mathcal{C}=&\,\frac{2^{4 \delta -1} \delta  (\pi G M^2)^\delta }{1-2 \delta } -\frac{(16 \pi )^{\delta }  G \left(G M^2\right)^{1+\delta}\delta  (\delta +1) (2 \delta -3) R_0}{3 (1-2 \delta )^2}  \nonumber \\
&+\mathcal{O}(R_0^2)\,.
\label{eq:C}
\end{align}
We observe that, for $\delta\rightarrow 1$ and $R_0\rightarrow 0$, we recover the standard expression $\mathcal{C}=-8\pi G M^2$ proper of Schwarzschild's BH.

Hence, the Tsallis entropy non-trivially affects the BH thermodynamic configuration, which might be either stable or unstable, depending on the value of the nonadditive parameter $\delta$. To find the analytical range of $\delta$ values allowing for a positive heat capacity, we expand Eq.~\eqref{eq:C} to the first order around $\delta=1$. By doing this, we get
\begin{widetext}
\begin{equation}
\label{HeatCap}
    \mathcal{C}\approx -8\pi G M^2 \left\{1-(\delta -1)\left[1-\ln(16)-\left(1-\frac{4}{3} G^2 M^2 R_0\right) \ln \left(\pi  G M^2\right)\right] -\frac{2}{3} G^2 M^2 R_0\left[11-4 \ln 4+\delta  (\ln (256)-9)\right] \right\}.
\end{equation}
\end{widetext}
Thus, requiring $C>0$ from Eq.~\eqref{HeatCap} yields
\begin{equation}
    \delta\lesssim 1\,-\, \frac{1}{\ln \left(16 \pi  G M^2\right)-1}\,+\,
    \frac{14 G^2 M^2 R_0}{3 \left[\ln \left(16 \pi  G M^2\right)-1\right]^2},
    \label{Deltacond}
\end{equation}
under the condition
\begin{equation}
\label{M}
    M\gtrsim M_0\equiv\frac{1}{4} \sqrt{\frac{e}{\pi G }} \,,
\end{equation}
where we have neglected the subdominant contributions in $R_0=4\Lambda\simeq 10^{-52}\,\text{m}^{-2}$ (in SI units). Indeed, regardless of the value of $M$ within the observational mass range, $M_\odot\lesssim M\lesssim10^{11} M_\odot$, the last term on the right-hand side of inequality \eqref{Deltacond} is negligible compared to the first two. We observe that solution~\eqref{Deltacond} represents the only physically viable bound. In fact, a closer inspection reveals that  $M_0\simeq5\times10^{-9}\,\mathrm{kg}$ (in SI units), making condition~\eqref{M} valid for any BH within the aforementioned mass range.

Some comments are in order here. First,  we emphasize that the Tsallis paradigm allows us to resolve the thermodynamic instability of Schwarzschild BHs for suitable values of the non-additive index. Furthermore, the upper bound on $\delta$ is slightly affected by variations in the BH mass, with smaller $M$ corresponding to a greater departure from the entropy
holographic scaling ($\delta = 1$). Specifically, considering the whole range of observed BH masses, one finds 
\begin{equation}
\label{dunmezzo}
    \delta\lesssim  0.995 \,.
\end{equation} 
Notably, albeit obtained within a different framework, this
constraint mainly overlaps with the range in Eq.~\eqref{CliftBar}.

\section{Conclusions}
\label{Concl}

We explored the gravitational origin of the Tsallis entropy, arising from a nonextensive modification of the BG statistics invoked to properly describe the properties of gravitational systems.
For our purpose, we considered BH solutions in the context of modified $f(R)$ theories of gravity, and we made use of Walds's formalism to deduce the Lagrangian associated with the Tsallis entropy. 

Starting from the $f(R)$ BH metric, we obtained the corrections to the Schwarzschild radius in the regime of small perturbations around GR. Subsequently, by computing the BH event horizon area and comparing the resulting entropy with the Tsallis expression, we found the analytic form of $f(R)$ as a function of the nonadditive parameter $\delta$ and an arbitrary constant. The latter was determined by requiring the matching of our result with the Schwarzschild-de Sitter solution in the standard entropy limit given by $\delta=1$.

The reconstructed effective gravitational Lagrangian was obtained by 
recasting $f(R)$ as a function of the dimensionless parameter $\epsilon$, which measures small departures from GR. In particular, we found $f(R)=c_1+c_2\left(1 + c_3 R\right)^{1+\epsilon}$,  representing a generalization of the well-known scenario $f(R)\propto R^{1+\epsilon}$~\cite{Clifton:2005aj}.
Thus, taking into account the constraints on $\epsilon$ derived from the abundances of light elements~\cite{Clifton:2005aj}, we discussed the physical implications of our result. Specifically, we inferred $0.983<\delta\lesssim1.001$, which aligns with recent predictions on Tsallis cosmology obtained from a combination of low and high-redshift observations~\cite{Asghari:2021lzu}. Additionally, we addressed the thermodynamic instability problem of Schwarzschild BHs, showing that the Tsallis entropy provides a potential solution under specific conditions. 
We would like to mention that corrections on the thermodynamic configuration of Schwarzschild's BHs have also been discussed in~\cite{El-Menoufi:2017kew}, showing that the stabilization might be prompted by quantum gravity effects. It would be insightful to connect such a result with our finding and possibly reinterpret the origin of the Tsallis entropy~\eqref{eq:Tsallis entropy} (or, equivalently, the modified gravity model~\eqref{eq:f(R)_final}) in the language of quantum theory. Our results call for a possible modification of GR on cosmological scales and within the context of BH physics in order to explain the observed phenomenology.

Finally, it is worth noting that the present analysis might be relevant beyond the specific framework under discussion. Indeed, in the working regime of $|\delta-1|\ll 1$, the Tsallis entropy takes the form of Eq.~\eqref{eq:S_new}. 
Logarithmic-like corrections to horizon scaling often appear in various quantum gravity theories, such as string theory, loop quantum gravity, the AdS/CFT correspondence, and the generalized uncertainty principle. In light of the close connections between gravity, cosmology, statistical and quantum mechanics, our research advances the comprehension of gravity and the description of spacetime from both the thermodynamic and information theory perspectives. Work along these directions is already underway and will be elaborated elsewhere.

\acknowledgments
{The authors are grateful to the anonymous Referee for useful comments on the manuscript.} 
R.D. acknowledges the financial support of INFN--Sezione di Napoli, \textit{iniziativa specifica} QGSKY. G.G.L. acknowledges the Spanish ``Ministerio de Universidades''  for the awarded Maria Zambrano fellowship and funding received from the European Union - NextGenerationEU.

\bibliography{references}

\end{document}